REVIEW ARTICLE

# The Life Cycle of Large Language Models: A Review of Biases in Education

**Jinsook Lee**[1] | **Yann Hicke**[2] | **Renzhe Yu**[3] | **Christopher Brooks**[4] | **René F. Kizilcec**[1]

[1]Information Science, Cornell University, New York, USA
[2]Computer Science, Cornell University, New York, USA
[3]Teachers College and Data Science Institute, Columbia University, New York, USA
[4]School of Information, University of Michigan, Ann Arbor, USA

**Correspondence**
Corresponding author Jinsook Lee.
Email: jl3369@cornell.edu

**Abstract**

Large Language Models (LLMs) are increasingly adopted in educational contexts to provide personalized support to students and teachers. The unprecedented capacity of LLM-based applications to understand and generate natural language can potentially improve instructional effectiveness and learning outcomes, but the integration of LLMs in education technology has renewed concerns over algorithmic bias which may exacerbate educational inequities. In this review, building on prior work on mapping the traditional machine learning life cycle, we provide a holistic map of the LLM life cycle from the initial development of LLMs to customizing pre-trained models for various applications in educational settings. We explain each step in the LLM life cycle and identify potential sources of bias that may arise in the context of education. We discuss why current measures of bias from traditional machine learning fail to transfer to LLM-generated content in education, such as tutoring conversations because the text is high-dimensional, there can be multiple correct responses, and tailoring responses may be pedagogically desirable rather than unfair. This review aims to clarify the complex nature of bias in LLM applications and provide practical guidance for their evaluation to promote educational equity.

**KEYWORDS**
Large Language Model (LLM), Bias and Fairness, Education

## 1 | INTRODUCTION

In late 2022, Large Language Models (LLMs) and generative artificial intelligence (AI) chatbots captured widespread attention when OpenAI released a public beta version of its LLM-based chatbot ChatGPT. It offered a compelling demonstration of the state of the art in generative AI chatbots by engaging in text-based conversations that exhibit forms of intelligence and a human-like tone. The technology was put to the test, quite literally, and scored extremely highly on a large variety of standardized tests, in addition to fooling a panel of judges in a version of the Turing test, which led scientists to question the validity of the famous benchmark for machine intelligence (Biever, 2023). Realizing the immense impact that LLMs can have in education, OpenAI partnered with Khan Academy ahead of the public release of GPT-4 to help the EdTech provider integrate a version of GPT-4 into its learning platform as an "AI-powered guide, tutor for learners, and assistant for teachers" called Khanmigo (Academy, n.d.). Similar AI-powered learning assistants quickly appeared in other major EdTech platforms, such as Coach on the Coursera platform (Coursera, 2023) and XPert on the EdX platform (edX Press, n.d.). These chatbots are perhaps the closest anyone has come to a scalable and domain-agnostic solution to Bloom's Two-Sigma Problem on how to provide large numbers of learners with support that is as effective as personal tutoring using a mastery-learning approach (Bloom, 1984). EdTech providers are developing new features using LLMs to enhance their products, including AI tutors that answer student questions in real-time, provide instant, personalized feedback on written assignments, or help teachers create new assignments and grade them faster with detailed feedback. There are numerous potential applications of this new technology in education (Yan et al., 2024), which raises questions about the long-term impacts of AI in education, and more immediate questions about issues that can arise when AI-based technology, built on data sourced from the World Wide Web, is deployed in classrooms (Yan et al., 2024; Denny et al., 2024).

In this review article, we focus on the potential biases that LLMs may exhibit in the context of education. Algorithmic biases tend to negatively impact members of disadvantaged groups and perpetuate inequities at a larger scale. Most LLMs, including GPT models (OpenAI, 2023; Bubeck et al., 2023; Brown et al., 2020; Radford et al., 2019), Palm 2 (Anil et al., 2023), BLOOM (Scao et al., 2022), LLaMA (Touvron et al., 2023), Flan-T5 (Chung et al., 2022), BERT (Devlin et al., 2018), RoBERTa(Y. Liu et al., 2019), are trained on extremely large web corpora, which can cause them to learn social biases even when active steps are taken to mitigate them. This can be difficult to examine directly because many LLMs, including those developed by OpenAI, are not released as open-source models and provide limited information on how models were trained and evaluated. A growing number of open models have been released, including Mistral (Jiang et al., 2023), Falcon Almazrouei et al. (2023), Gemma (G. Team et al., 2024), Gemini (G. Team et al., 2023) and QWen-7B (J. Bai et al., 2023)), which offer more insights into the model's fairness properties. Still, biases can also arise based on how models are integrated into an application, which has sparked efforts to promote responsible AI using application-specific licensing (e.g., the BigScience RAIL License[†]).

The rapid adoption of LLM-based technology in educational institutions presses the need to systematically evaluate LLMs for bias to avoid unintended consequences, such as amplifying current educational inequities in opportunity and achievement. Although there is an established area of research on AI bias and fairness, including a domain-specific literature for education (Baker & Hawn, 2021; Kizilcec & Lee, 2022), there is limited guidance on what potential biases can arise in the process of LLM development, how to evaluate and mitigate bias in LLM-based applications, specifically in the context of education. Applications of LLM-based generative AI raise particular challenges for evaluating bias due to the complexity of its natural language output and establishing a ground truth that is appropriate for the context of use. This review aims to improve our understanding of bias resulting from LLMs in educational applications. To define the context of these applications, we first review a set of studies that use LLM technology to support a variety of tasks in educational settings. Then, building on an established life cycle framework of (traditional) machine learning, we propose a model of the LLM life cycle that traces each step from the initial development to the final touches of customization for LLM-based applications. For each step in the LLM life cycle, we review potential biases that can arise in educational contexts and potential measures of those biases. We discuss the implications of the LLM life cycle for researchers interested in evaluating and mitigating bias, practitioners interested in understanding where biases might arise from, and policymakers looking to better understand the ethical issues related to LLM use in education. This review highlights opportunities and practical challenges of using LLMs in education and important areas for future research on LLM bias and fairness in education.

## 2 | LLM APPLICATIONS IN EDUCATION

There are a variety of ways that LLMs can be used in educational contexts, many of which have been described by Yan et al. (2024). We organize them into two broad types of use cases: **natural language generation** (NLG) and **natural language understanding** (NLU) tasks. NLG tasks include creating educational content, such as lesson plans, assessments, and in-class materials like worksheets (Kasneci et al., 2023; Wollny et al., 2021; Leiker et al., 2023). NLU tasks involve analyzing text for an educational purpose, such as making a prediction based on a student's essay submission about how well they understood the materials and scored on a given grading rubric. NLU tasks can also serve as an input into a larger model, such as an LLM used to detect confusion in a student's question, which can serve as an input into a predictive model for student underperformance and drop-out. An NLU task can also serve as the first step of an NLG task: an AI-based grading system, for example, may first analyze and score a student's essay and then generate written feedback based on that analysis (L. Zheng et al., 2022). Other examples of combined NLU-NLG tasks include tutoring chatbots like Khanmigo (Academy, n.d.) and Rori (Henkel et al., 2024), which provide customized guidance to students across subjects including mathematics and the language arts, systems that provide personalized hints for compiler errors in a programming course (Pankiewicz & Baker, 2024), and tools designed to provide feedback or training to educators and tutors (J. Lin et al., 2023). In the context of this review article, we focus on cases where LLMs are used to enhance teaching and learning, and we therefore do not consider use cases like LLM-based essay-writing services.

In the responsible AI literature, algorithmic biases have been organized into two broad categories representational biases and allocative biases (Suresh & Guttag, 2021). The potential biases associated with NLG tasks are mostly representational biases because NLG tasks can create text containing stereotypes or misrepresentations, exclusionary language, or even toxic content

[†] https://huggingface.co/spaces/bigscience/license

(Weidinger et al., 2021).[‡] The potential biases associated with NLU tasks are mostly allocative biases because individuals may receive differential access to resources or opportunities (Suresh & Guttag, 2021). For example, a grading system using an LLM for NLU could systematically assign lower scores to students of certain demographic groups, even though no identifying information was provided to the LLM. In fact, LLMs have been shown to display dialect prejudice when asked to make decisions about speakers of African American English (AAE) as compared to speakers of Standard American English (SAE) (Hofmann et al., 2024). Educational applications that rely on both NLU and NLG are susceptible to both types of biases. For example, an intelligent tutoring system might generate assessments that inadvertently reinforce stereotypes (a representational bias) and also disproportionately show those assessments to students with certain backgrounds (an allocative bias).

The classification of different tasks (NLU and NLG) and types of biases (allocative and representational) begin to organize the complexity associated with bias from LLMs in education technology. However, it does not explain where biases originate in the multi-step process from developing to customizing to ultimately deploying an LLM for education. We therefore developed a detailed model of this multi-step process that can help identify where biases might emerge, for what reasons, and how to potentially measure them.

## 3 | THE LLM LIFE CYCLE FROM DEVELOPMENT TO DEPLOYMENT

We build on the framework of the machine learning life cycle proposed by Suresh & Guttag (2021). It pinpoints where bias can be introduced in the process of creating and deploying a system using traditional machine learning. We have modified their original framework for the specific context of LLM-based applications, which is substantially more complex, to examine where bias may be introduced (Figures 1 and 2). Due to its complexity, we divide the life cycle into two phases: the initial development phase of the base LLM, and the customization phase which relies on a base LLM. We describe potential biases in each step of the life cycle with examples from education contexts.

### 3.1 | Phase 1: Training a Base LLM

#### 3.1.1 | Scraping and Sampling

Large Language Models (LLMs) are trained using extensive text corpora, such as WebText or Common Crawl (Radford et al., 2019), which are scraped from pages on the World Wide Web. Online text data can reflect both current and past discrimination. Biases can arise from prejudices contained in these data, including biases inherent in the text (i.e. the content of the text) or biases arising from the selection process (i.e. which texts are included and which are excluded). **Historical bias** frequently arises when data is collected over a long period and unintentionally reveals historical discrimination for certain groups. For instance, when collecting data related to STEM fields, there tends to be an imbalanced gender representation because there has historically been less representation of women in these areas. Additionally, due to the vast amount of data from various sources, genres, and periods, the content may include discriminatory elements, such as documents involved in discrimination, which can pose harm to certain groups (Barocas & Selbst, 2016).

Considering the historical biases that have accumulated globally, **representation bias** can emerge in the form of an imbalance in the sampled data along dimensions including language, sample periods, available sources, and authorship. Ultimately, there is no way to avoid these difficult choices during the sampling process to narrow down the vast volume and diversity of text on the Internet. Representation bias can arise due to source availability and related policy restrictions, resulting in a predominant collection of English-focused datasets, while datasets for other languages could be relatively underrepresented. Consequently, content in other languages might not be fully represented in the actual world. Additionally, the choice of when to start scraping and sampling can cause representation bias because data gathered a long time ago might not reflect present-day conditions. The consequences of representation bias, including geographical (Ocumpaugh et al., 2014) and temporal (Levin et al., 2022) bias in the training data, have been examined in the context of education technology. Yet representation bias can occur not only during data sampling but also when recruiting people for data labeling or "red teaming" (the practice of recruiting an external team to discover risks by taking an adversarial approach, for example, showing biases by trying to elicit them from the system). The background characteristics of individuals recruited for these efforts can present a further source of representation bias.

[‡] In subsequent sections, we will be using the term "representation bias" to refer to both what Suresh and colleagues (Suresh & Guttag, 2021) refer to as representational bias (stereotypes, misrepresentation, toxic content, etc.) and also imbalances in training or fine-tuning datasets compared to a target population (e.g., an underrepresentation of women-authored texts), which is typically referred to as a representation bias. While these are distinct concepts, it is clear in a context which one is relevant, and so we opted for the simpler presentation by using one of the two phrases throughout.

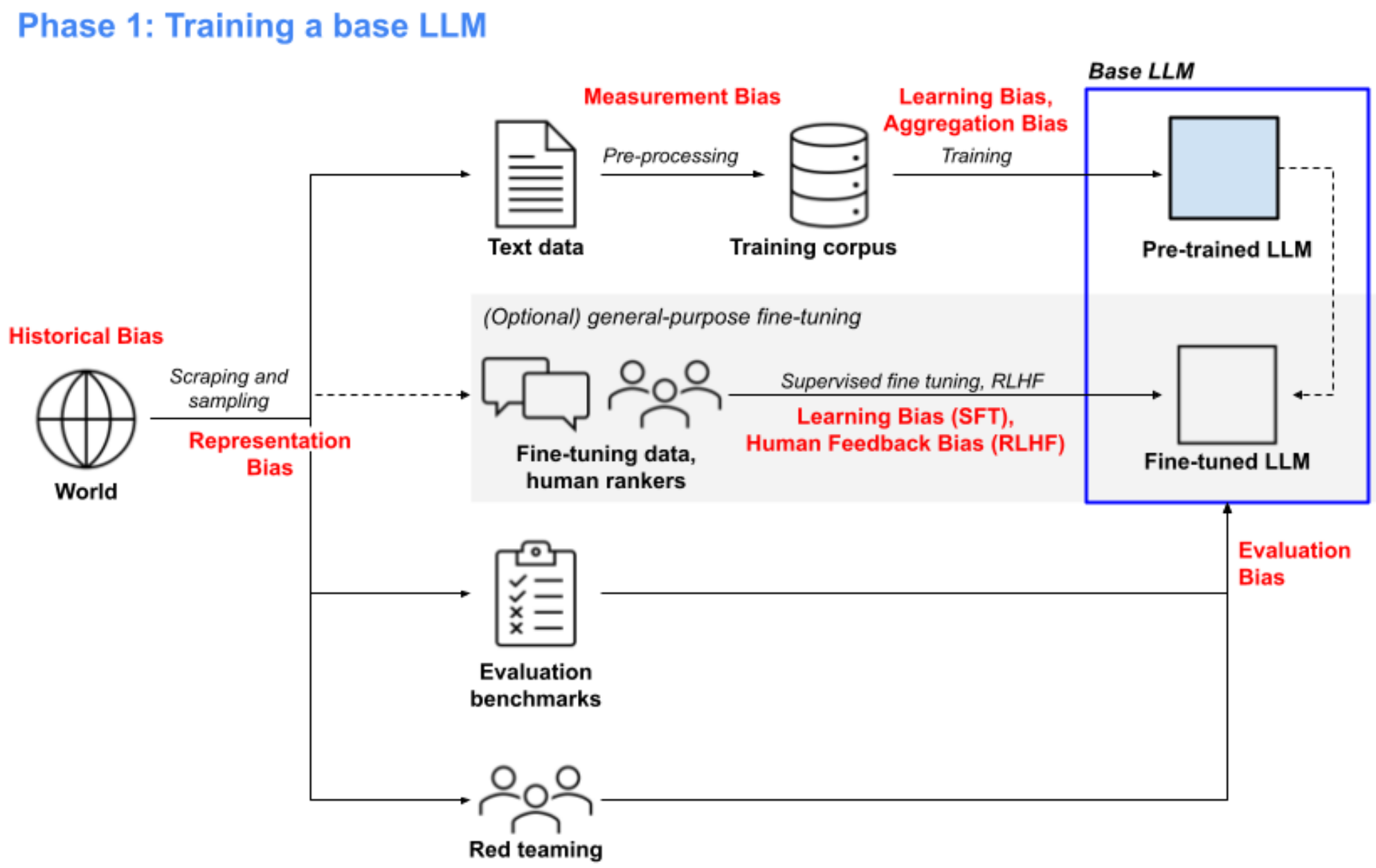

**FIGURE 1** The Initial Development Phase of the LLM Life Cycle with Potential Sources of Biases, after Suresh & Guttag (2021)

The unregulated nature of World Wide Web content can further contribute to representation bias. Specifically, *harmful content* that is explicitly or implicitly stereotyping, misrepresenting, and using toxic or exclusionary language can affect the representation of members of certain groups in the training corpus. A number of open training datasets, such as *LAION-400M* (T. L. Team, 2021), have been found to contain disturbing and explicit content, including images-text pairs related to rape, pornography, harmful stereotypes, as well as racist and derogatory remarks about some ethnic backgrounds (Birhane et al., 2024, 2021). This evidence suggests that larger-scale versions of these datasets could exacerbate representational bias.

Once a text corpus for developing the LLM has been sampled, the next step is to *pre-process* the data. To improve data quality, duplicate texts are removed, noisy data are removed (e.g., very short pieces of text), personally identifiable information is removed or masked, text related to popular benchmarks is removed to ensure a fair evaluation (a process known as decontamination), and texts containing toxic or overtly biased language are removed (Weights & Biases, 2023). The process of filtering out toxic and biased content relies on dictionaries (LDNOOBW, 2023) or detection tools (spamscanner, 2023). However, these may not capture all instances of objectionable speech. We can apply a framework to help parse harmful content, for instance, by categorizing it along the type of harm (e.g., misinformation, hate speech, stereotypes), whether harmful content is sought out for the specific application (e.g., to learn how to identify it better going forward) or not, and who is affected by the harmful content (e.g., individuals represented in the dataset, demographic groups) (Kirk et al., 2022).

In developing tools or frameworks to process raw text data, we may inadvertently encounter **measurement bias**, defined here as a systematic error in measuring specific abstract concepts (e.g., toxicity, bias, private information). A feature typically represents a specific measurement that stands in for a broader and often intangible concept. For example, it can be challenging to measure the concept of "toxic" when there are only subtle and implicit discriminatory words in the text. If certain slang terms are commonly used by a particular community, it can be difficult to determine whether the words are toxic or not. Measurement bias can also arise from people tasked with identifying instances of the construct. The opinions of individuals who label toxic and biased content are shaped by their viewpoint and background, which can reinforce their perspectives (and exclude others) through the process of data curation and filtering (Weights & Biases, 2023). Overall, this inherent ambiguity in quantifying abstract constructs can introduce measurement bias when operationalizing these constructs during data pre-processing, and ultimately lead to harm (Jacobs & Wallach, 2021).

### 3.1.2 | Pre-training (Training Corpus → Pre-trained LLM)

Once the training corpus is pre-processed, the next step is *tokenization*. The text data is broken down into pieces that can be words, parts of words, or byte pairs. This process transforms the text corpus into a format that models can process. These are used to create word and contextual embeddings to represent features that allow machine learning models to easily correlate input data with output data. These embeddings are designed to capture the semantic and syntactic properties of words within a high-dimensional space, thereby enhancing the model's capability in NLU and NLG tasks. This sets up the architecture for pre-training the model, which involves a sequence of transformer blocks with multi-head self-attention mechanisms and fully connected layers of neural networks (Radford et al., 2018).

Once this architecture is set up, the model is pre-trained to predict the next token in sequence, and during the pre-training, the model's weights are optimized while its predictions are continuously compared to the actual outcomes, using the errors to update the weights in each step. The model thereby learns contextualized representations of words and phrases. Typically, the loss function used for pre-training LLMs is cross-entropy loss, which measures the difference between the predicted probability distribution and the distribution of the actual next token (Mehrabi et al., 2021; Minaee et al., 2024). Pre-training techniques vary based on whether the focus is on NLU or NLG. For NLU, models like BERT utilize masking techniques where some words in a sentence are hidden, and the model is trained to predict these masked words. This approach helps the model grasp the context and meaning of sentences. Additionally, BERT employs a next-sentence prediction task where the model predicts whether a sentence logically follows a given sentence, further enhancing its understanding capabilities. On the other hand, NLG-focused models like GPT are pre-trained using a next-word prediction task, where the model learns to predict the next word in a sequence given the previous words. This sequential prediction task is important for generating coherent and contextually relevant text (Solaiman & Dennison, 2021).

However, **learning bias** (also known as algorithmic bias) can arise during this process driven by an objective function like minimizing cross-entropy loss if undesirable biases in the training data are inadvertently amplified. We define learning bias in LLMs as *amplifying undesirable inherent biases* when there is a goal to minimize a given loss function. The bias that is encoded in this step can be considered intrinsic to the model because it resides in the geometry of the embedding space (Goldfarb-Tarrant et al., 2020). There is a plethora of studies examining learning bias in word and contextual embedding spaces, including ones that study gender bias (Bolukbasi et al., 2016; J. Zhao et al., 2019), gender and ethnic stereotypes (N. Garg et al., 2018), gender neural words (J. Zhao et al., 2019), cultural biases (Tao et al., 2023; Durrheim et al., 2023), and studies that trace training documents to identify the origin of such biases (Brunet et al., 2019). There is also extensive research on debiasing word embeddings, such as reducing gender bias (Bolukbasi et al., 2016; Gonen & Goldberg, 2019).

Intrinsic learning bias can be measured with either embedding-based metrics or probability-based metrics. Embedding-based metrics are computed distances in the vector space between words/sentences representing the domain of evaluation (e.g., professions) and words/sentences representing the identities being evaluated for bias (e.g., genders, racial groups). The Word Embedding Association Test (WEAT) Caliskan et al. (2017) is a commonly used embedding-based metric that quantifies biases in word embedding by examining how closely words related to certain concepts are associated with words related to social groups or attributes. Likewise, the Sentence Encoder Association Test (SEAT) quantifies bias in a set of sentences by encoding them into numerical embeddings using a sentence encoder model (May et al., 2019). Probability-based metrics are computed based on the likelihood of predictions. For example, the Discovery of Correlations (DisCo) method Webster et al. (2020) uses masked tokens in a template sentence completion task. The first part of the template sentence includes a word related to a specific social group (e.g., gendered names or pronouns), and the second part has the language model predict the top three words that might complete the sentence. DisCo counts how often the model predicts different words for different social groups across all templates to obtain a probability-based measure of bias. While DisCo focuses on uncovering patterns within the model's predictions, the Log-Probability Bias Score (LPBS) (Kurita et al., 2019) measures intrinsic probability distributions of the model's outputs by directly measuring how likely the model is to produce certain biased outputs based on the log-probabilities.

Finally, **aggregation bias** can arise when a chosen model does not perform equally well across all subgroups, often because the data includes distinct subgroups that are treated uniformly instead of individually (Hutiri & Ding, 2022). There may not be a one-size-fits-all model that does not make any sacrifice on performance for certain groups. This bias is relevant for both NLU and NLG tasks, for example, in that a model works well for one language but is not the optimal choice for other languages.

### 3.1.3 | (Optional) General-purpose Fine-tuning

After pre-training the language model, LLM developers may use general-purpose fine-tuning to adjust their model to improve its performance across a wide range of tasks instead of optimizing it for a specific task. This can be achieved using supervised fine-tuning (SFT) to adapt the model parameters to behave in a certain way using a specific dataset and a supervised target (Radford et al., 2018). Since this step adds an additional dataset, **representation biases** can be introduced here too. And since SFT updates the pre-trained model parameters based on a chosen objective function, **learning biases** can arise as well.

Reinforcement Learning from Human Feedback (RLHF), which is a type of SFT, is increasingly used to fine-tune the model's behavior to better align with the goals, needs, or preferences of a user group (OpenAI, 2023). Human raters are recruited to provide a large number of rankings of text outputs based on criteria such as harmlessness and helpfulness (Y. Bai, Kadavath, et al., 2022). The resulting dataset contains important signals for what output is more desirable for a particular user group, a domain, or a task, but **human feedback bias** can be introduced in this step. Human feedback bias creates issues when these ratings mistakenly reinforce a model to behave in undesirable ways. The RLHF process requires high-quality feedback data, and undesirable outcomes can occur if the instructions provided during the labeling process are insufficient or unclear. For example, without proper guidance and training, human raters might generate preference data that leads the model to suggest harmful actions, such as criminal activity (OpenAI, 2023).

Beyond human feedback bias, **representation bias**, **measurement bias**, and **learning bias** can also emerge during RLHF. Representation bias can arise if the sample characteristics of the human raters do not adequately represent the relevant population of the model's application context. Measurement bias can arise because concepts like harmlessness and helpfulness are abstract, and human raters might have varying standards in mind when making judgments. Learning bias can occur during the process of updating model parameters, depending on how the reward model is created and the objective function is chosen. The challenges and open problems associated with human feedback bias in RLHF include that human raters may pursue incorrect and harmful goals, including giving adversarial ratings that are hard to spot but that can lead to data poisoning (Casper et al., 2023).

### 3.1.4 | Base LLM Evaluation

Before the base LLM is ready, it needs to undergo an evaluation step. Many benchmark datasets have been created for the purpose of evaluating LLMs by testing different aspects of the model's capacities on NLU and NLG tasks. In addition to standard benchmark datasets, some developers may engage a group of critical external testers to find flaws or vulnerabilities in a model's performance and behavior–a process known as red teaming. This adversarial approach helps uncover potential weaknesses that might not be evident through standard evaluation methods. By actively trying to break the model or cause it to produce incorrect or biased outputs, the red teaming provides valuable insights into the model's robustness and safety (Ganguli et al., 2022). A popular framework for evaluating LLMs is the Holistic Evaluation of Language Models (HELM) project (Liang et al., 2022), which includes a number of evaluations that focus on the interpretability and transparency of models, including bias metrics such as toxicity.

**Evaluation bias** can arise in this step because there are many choices for evaluating the model, which can lead to substantially different conclusions. First, since benchmark datasets are also scraped and sampled from available sources on the Internet, they may fail to represent all relevant user groups, and **historical bias** and **representation bias** can emerge. Additionally, if the benchmark datasets contain construct measures that fail to serve as a valid "proxy", **measurement bias** can emerge. Second, the composition of a red team can be biased and skew the evaluation results. Likewise, although machine learning researchers often have access to many statistical methods and models, they tend to select only a few results to report based on their personal preferences and available resources (Young, 2018). This selective reporting can create a "garden of forking paths" (Young, 2018), where different choices in the analysis process lead to significantly different and potentially incorrect results. This issue also arises in the development of LLMs. It underscores the importance of considering model uncertainty during the evaluation step to enhance the credibility and reliability of the models, especially given the growing skepticism and concerns about the potential harm from biased or incorrect outputs. This is crucial because the choice of performance metric, benchmark dataset, and red teaming approach can all influence the evaluation results.

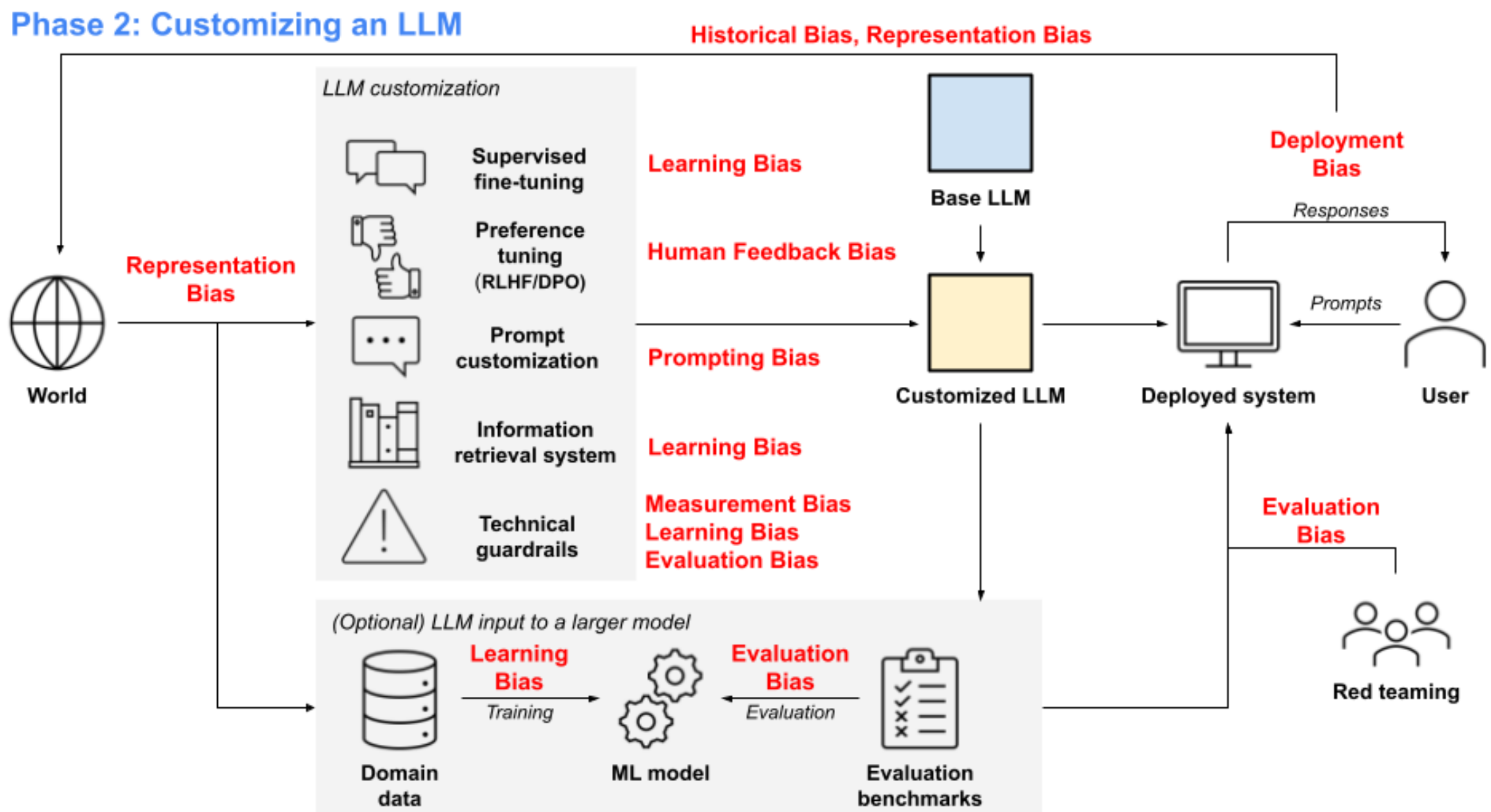

**FIGURE 2** The Customization Phase of the LLM Life Cycle with Potential Sources of Biases, after Suresh & Guttag (2021)

## 3.2 | Phase 2: Customizing an LLM

### 3.2.1 | LLM Customization

After the base model is evaluated, education practitioners can tailor the model for their specific needs using various customization techniques (Figure 2). A popular technique for customizing an LLM is SFT (**supervised fine-tuning**), which refines the base model using a dataset to specialize the model in a particular domain or task (X. Liu et al., 2021; H. Zheng et al., 2023). For instance, *FineWeb-Edu* (Lozhkov et al., 2024) is an education-specific dataset (derived from the CommonCrawl dataset) comprising 1.3 trillion tokens for use in LLM customization. The resulting fine-tuned model retains the extensive knowledge embedded in the base model and additionally incorporates domain-specific information. In the education context, this method has been applied to improve automatic assessment scoring (Latif & Zhai, 2024), to support math tutors for remediation of students' mistakes (R. E. Wang et al., 2023), to assess personal qualities in college admission essays (Lira et al., 2023), and to reduce performance disparities in math problem skill tagging tasks across different languages (Kwak & Pardos, 2024).

Another potential technique for customizing an LLM is **preference tuning** using RLHF or Direct Preference Optimization (DPO). DPO is a fine-tuning method inspired by reinforcement learning that is relatively simple, stable, and computationally efficient; it outperforms commonly used methods such as Proximal Policy Optimization (PPO) based RLHF in many cases (Rafailov et al., 2024). DPO leverages the relationship between the reward model and optimal policies, efficiently addressing the challenge of constrained reward optimization within a single policy training phase using human preference data. Both SFT and DPO techniques have been applied in education, for example, to create an intelligent question-answering system that is tailored to a specific introductory computer science course (Hicke et al., 2023).

When practitioners or researchers fine-tune a base model, their domain- or task-specific dataset and any human preference data they collect are vulnerable to both **historical bias** and **representation bias**. These datasets are often sourced from the Internet, smaller in size, and focused narrowly on specialized tasks that reflect the characteristics of the domain. If this dataset mirrors skewed societal perspectives or inaccuracies or represents only a specific group of people, the fine-tuned model might adopt these biases, making it less generalizable and more likely to make prejudiced decisions. Additionally, even if the dataset fairly represents the real world, **learning bias** can arise. In addition to amplifying undesirable biases in the training data, the model might overly adapt to the new dataset while updating parameters and forgetting some of the broader generalizations it had learned. This phenomenon is known as "catastrophic forgetting" (French, 1999) where domain-specific data overrides essential general knowledge. In the context of LLMs, Luo et al. (2023) conducted an empirical investigation and discovered that catastrophic forgetting is prevalent when fine-tuning LLMs such as Llama-7b and Alpaca-7B. Additionally, Zhai et al. (2023) found that fine-tuning multi-modal LLMs can lead to increased hallucinations. In the context of education, this could hypothetically mean that fine-tuning a base model on mathematics textbooks, for instance, could overly specialize the model in mathematics and cause it to give less helpful or even inaccurate responses in other subject areas.

The domain-specific dataset used for fine-tuning is typically collected either from a platform the LLM developers had already created or surveys, which can give rise to **measurement bias**. Measurement bias has been studied extensively in educational data, which commonly has a nested, multilevel structure because students are observed within classrooms, or each student might be given a different subset of questions for a standardized test (Jak et al., 2014). This type of measurement bias can be detected using structural equation modeling (SEM) with respect to different attributes, including student demographics, teacher demographics, and classroom characteristics. Another example of measurement bias can arise from unexpected (and possibly unobserved) patterns in the data collection process. For example, Ogan et al. (2012) examined how an intelligent tutoring system (ITS) was used in classrooms in Latin America and found that many students worked collaboratively to solve problems, even though the system was designed for individual use. This can create measurement bias in ITS data that could be used for fine-tuning an LLM because high performance might be inaccurately attributed to individual students when they were actually collaborating.

If direct access to the fine-tuned LLM's internal parameters or embeddings is available, learning bias resulting from fine-tuning can be measured using the same embedding- and probability-based measures described above. Alternatively, extrinsic bias measures that systematically evaluate text generated by a fine-tuned model in response to specific prompts can be used (Delobelle et al., 2022). For this output evaluation, the token distribution between different social groups is compared using distribution metrics, classifier metrics, or lexicon metrics. Distribution metrics compare the distribution of explicit or implicit mentions of social groups to a baseline distribution (Bommasani et al., 2023). These metrics compare differences in the percentage of predictions that exactly match the ground truth (i.e. exact match) (Rajpurkar et al., 2016), or use co-occurrence measures (Bordia & Bowman, 2019) to detect variance in group representation. For example, the Perspective API (Google Jigsaw, Year of Access) measures toxicity by providing a toxicity probability for generated text. Sicilia & Alikhani (2023) suggested using Score Parity to assess how consistently a language model generates text based on certain attributes (e.g., toxicity) across different protected attributes (e.g., demographic groups). Lexicon-based metrics parse generated text at the word level, comparing words to pre-defined lists of harmful or biased terms, and assigning predefined bias scores to each word. Examples of lexicon-based metrics include HONEST (Nozza et al., 2021), which measures harmful words in generated text, and BOLD (Dhamala et al., 2021), which measures psycho-linguistic norms by assigning affective values (e.g., dominance, sadness) to words and calculating text-level norms as weighted averages; Gender Polarity (Dhamala et al., 2021) measures the frequency of gendered words in the generated text.

Another technique for LLM customization is **prompt customization** (also called "prompt tuning"). This method adapts pre-trained transformers to specific tasks by modifying the input prompts rather than changing the model's internal parameters (P. Liu et al., 2023). It leverages the inherent knowledge within pre-trained models to enhance its task-specific performance. Optimized prompt-tuning can be as effective as fine-tuning across models of various sizes and across different tasks (X. Liu et al., 2021). However, this technique can give rise to at least three types of **prompting bias**: majority label bias, recency bias (overemphasizing the importance of the latest information), and common token bias (Z. Zhao et al., 2021). These can cause pre-trained LLMs to exhibit **representation bias** towards specific responses: for instance, if the final response prompt contains a negative label, it may influence the model to predict negative language. To measure prompting bias, Kotek et al. (2023) proposed a paradigm to test gender bias in LLMs by using a set of 15 prompts that contain stereotyping contexts to evaluate the susceptibility of a model.

Finally, a developer might use information retrieval in the LLM life cycle to generate responses grounded in information from relevant data sources, applying a customization technique called **Retrieval-Augmented Generation** (RAG) (P. Lewis et al., 2020). This technique allows the model to refer to external information for generating responses using two primary types of retrieval: knowledge-based and API-based retrieval. Knowledge-based retrieval systems store the current context in a "vector store," an embedding space where users can query and find related content similar to the query. This ensures the LLM remains up-to-date and contextually relevant to specific downstream tasks. For example, a course-specific educational chatbot might use RAG to answer students' questions based on the official course materials (Hicke et al., 2023). API-based retrieval uses external databases, such as learning management systems (LMS) or student enrollment databases, to generate responses. This provides additional context, enhancing the quality of responses to be more personalized and relevant. However, combining LLMs with external databases to provide better context can introduce **representation bias** and **measurement bias** because of how contextual data is archived or integrated. Additionally, the retrieval system often uses ranking algorithms to sort the most similar contexts based on the user's query. These ranking systems can introduce **learning bias**, as they may favor certain types of content over others, possibly reducing the level of diversity in the set of retrieved documents.

Most LLM developers carefully review and monitor their base models and (to the extent possible) their customized models for the potential biases described above. They have formalized these checks into a set of **technical guardrails** for LLMs, which are critical frameworks and procedures aimed at promoting the ethical, secure, and accountable deployment of LLMs. These guardrails include content filters to prevent the generation of harmful or inappropriate content, usage monitoring to detect and mitigate misuse, and model tuning to reduce biases and enhance fairness. Additionally, technical guardrails may involve implementing privacy-preserving techniques to protect user data and incorporating explainability features to make the model's decisions more transparent (Attri, 2023). For example, Meta's Llama Guard (Inan et al., 2023) provides a holistic and thorough evaluation framework for responsible LLMs, while acknowledging its limitations, including that it focused on English, which can cause **representation bias** in other languages. Once again, as with any customization that attempts to measure and mitigate issues (e.g., applying toxicity classification), the ML approach implemented within the technical guardrail is susceptible to **measurement bias** and **learning bias**.

### 3.2.2 | Deployment (Customized LLM → Deployed System)

Once the customization is finalized and the quality of outputs is tested, the model can be deployed in various forms based on whether the model is solving NLU, NLG, or combined NLU-NLG tasks. When deploying a customized model such as Khanmigo (Academy, n.d.) or Rori (Henkel et al., 2024), there can be a gap between the problem that the system was originally designed to address and the way it is used in practice, a so-called **deployment bias**. Deployment bias can be observed as a difference in application usage and performance across populations (Gallegos et al., 2023). For example, the study of students in South America using an ITS collaboratively, instead of individually as intended by the system designer, exemplifies the importance of deployment bias in authentic educational contexts (Ogan et al., 2012).

The deployment step in the LLM life cycle is particularly important because of the potential harm arising from human-computer interaction. When LLMs are used as "conversational agents" (Perez-Marin & Pascual-Nieto, 2011), they can "speak" in natural language, a primary mode of human communication. As a result, users might anthropomorphize these systems, viewing them as human-like, which can lead to overreliance or unsafe use. This presents a critical issue in education. For instance, if students over rely on an ITS because it appears adept at generating empathetic and expert responses, they may place undue trust in potentially unethical, unverified, or hallucinated information it generates. Students could be misled and engage in irresponsible academic practices. Another issue to consider with deployed LLM-based systems is how they might influence people's communication patterns. AI has been found to enhance communication efficiency and positive emotional expression, leading to closer and more cooperative interpersonal perceptions, but its use in conversations can be socially stigmatized and result in social harm (Hohenstein et al., 2023).

Finally, as LLMs are increasingly used to generate text, it is inevitable that text corpora used for training and customizing future models will include significant amounts of text generated by previous LLMs, rather than human authors. This may inadvertently amplify **historical and representation biases** that remained unaddressed in current LLMs (A. Wang et al., 2024), or biases that arise from who is predominantly generating text using LLMs and for what purposes. The continued induction of LLM-generated text into the population of all texts in the world will "pollute" datasets that represent human language, but it will also reflect the continuously evolving nature of language.

### 3.2.3 | (Optional) LLM Input to a Larger Model

Customized LLMs can be utilized by taking their outputs as inputs for more extensive machine learning frameworks. For example, an LLM can be used to assess students learning behaviors and status (e.g., whether they experience confusion or have misconceptions that prevent them from solving a problem) (H. Li et al., 2024). Combining the NLU and NLG capabilities of LLMs can enable applications such as automated essay grading systems, which evaluate essays and give natural language feedback on aspects including statistical measures (e.g., length and sentence complexity), stylistic elements (e.g., syntax, grammar, and punctuation), and content quality (e.g., accuracy, coherence, and key concept articulation) (Ramesh & Sanampudi, 2022). The prevalence of AI-generated content that could potentially influence student writing raises an emerging consideration about the ethics of AI plagiarism. For instance, LLMs can be used in an automated essay grading system to extract key features from student essays (NLU) and produce synthetic text (NLG) that can be used to check for originality against the student's work.

## 4 | DISCUSSION

The life cycle of traditional machine learning applications, which focus on predicting labels, is well understood (Suresh & Guttag, 2021). Biases are known to enter at various points in this life cycle, and methods to measure and mitigate these biases have been developed and tested, including in the context of education (Baker & Hawn, 2021; Kizilcec & Lee, 2022). However, with the increasing adoption of LLMs and other forms of generative AI in education, current evaluation approaches do not adequately address needs specific to supporting educational goals (Yan et al., 2024; Denny et al., 2024). This review contributes a holistic perspective of the LLM life cycle, using domain-specific examples in education to highlight opportunities and challenges for incorporating NLU and NLG supports into education technology applications. We identify the potential sources of bias at each step of the LLM life cycle and discuss them in the context of education. This offers a framework for understanding where potential harms of LLMs might arise for students, teachers, and other users of generative AI technology in education, which can guide approaches to bias measurement and mitigation.

Considering the important role of language in teaching and learning, LLMs are inevitably going to be a part of AI-based educational decision support systems (AI-EDSS). For educational practitioners and policymakers, it is crucial to be well aware of the types of biases that can originate from various steps in the LLM life cycle. The life cycle perspective can offer them a heuristic for asking technology developers to explain each step to help them assess the risk of bias and potential harm. We have argued that measuring the biases in systems that use LLMs is more complex than in traditional ML, primarily because evaluating NLG is highly context-dependent; what constitutes good feedback on a homework assignment, for instance, can vary widely. Education technology developers can play a significant role in collecting and curating datasets for LLM evaluation and benchmarks, which can be combined with collections of educational content scraped from the Internet and filtered for quality, such as the *FineWeb-Edu* dataset (**?**). To further tailor LLM-based systems for specific educational applications, participatory design methods to quickly prototype and collect feedback have been shown to work well in a holistic, evaluation-driven design approach (e.g., the LearnLM project between Google and Arizona State University; Irina Jurenka & et al. (2024).

A fundamental challenge for most evaluation protocols and objective functions used with LLMs is the use of short feedback cycles, which is especially problematic in educational contexts that aim to support students' long-term growth as critical thinkers and problem solvers. This contrast is exemplified by the fact that most current evaluation methods examine just an isolated model output (single-turn), instead of an entire conversation (multi-turn), let alone evaluating how a conversation impacts future opportunities to demonstrate a deeper understanding of the topic. Recent work using NLU to parse tutoring conversations to provide teachers with targeted feedback shows promise by moving towards a more holistic evaluation of multi-turn conversations (R. E. Wang & Demszky, 2024; Demszky et al., 2023). As the popularization of LLMs around 2023 happens to coincide with a broader movement in education to expand direct tutoring offerings (Loeb et al., 2023), it raises important questions about the effectiveness and responsible use of LLMs for on-demand tutoring, including as part of an ITS (D'Mello & Graesser, 2023).

In this review, we examined potential biases at each step in the life cycle of LLMs deployed in educational contexts. Our goal was to provide a clear and holistic review of the various sources of bias. In the process, we discovered that the same type of bias (e.g., representation, learning, and measurement bias) can arise in multiple steps of the process, in part because the LLM life cycle is more complex than traditional machine learning. This means that a carefully de-biased base model may get tainted (unwittingly) by a biased dataset introduced during model customization. Checking for bias in the final deployed LLM is arguably the most important step in assessing potential harms. The life cycle perspective can then help with identifying where observed biases may have been introduced. We describe relevant measures for assessing bias (and fairness) for NLG tasks with LLMs to guide practitioners and developers in the process of evaluation. Most measures evaluate the uniformity of outputs across social groups or detect the prevalence of group-specific biases (Gallegos et al., 2023; Liang et al., 2022). This approach raises an important question about how to negotiate the space between two desirable but ostensibly incompatible properties of LLM applications: fairness, which demands that similar queries/students receive the same responses, and personalization, which encourages responses to be non-generic and tailored to students' needs. We present this tension as a topic for future research, which could examine how LLMs should behave when there is no single correct response, and group-based adaptation may be pedagogically appropriate.

We conclude with three recommendations for future research on LLMs in education. First, there is a need for education-focused benchmark datasets that better represent a broader range of sociodemographic groups across the world, especially considering that applications like Khanmigo are expected to be used by a diverse group of students and teachers (Gallegos et al., 2023). Additionally, there is a need for high-quality education datasets for pre-training and fine-tuning models (Y. Li et al., 2023; Kwak & Pardos, 2024; Lozhkov et al., 2024). Second, there is a need to develop a specific taxonomy of harms for LLMs in educational contexts that promote responsible use and highlight the perspectives of educators, students, and their families. Current taxonomies tend to be domain-agnostic and developer-centered (e.g., Weidinger et al. (2022)). Finally, there appears to be a significant opportunity to use high-quality human feedback from multi-turn scenarios to improve the efficacy and alignment of LLMs with educational objectives (Chung et al., 2022; Zhou et al., 2024), for instance, to refine them for specialized tasks such as math tutoring (Irina Jurenka & et al., 2024).

### FUNDING INFORMATION

This work has been supported by funding from [omitted for review].

### CONFLICT OF INTEREST

There is no potential conflict of interest in this work.